%% file: crowd7.tex
\documentclass[11pt,preprint]{aastex62}

\usepackage{natbib}
\usepackage{float}
\usepackage{graphicx}
\usepackage{amssymb,amsmath,mathtools}

\received{}
\accepted{}
\shorttitle{Cepheid Amplitudes}
\shortauthors{Riess et al.}

\newcommand{\xddots}{%
  \raise 5pt \hbox {.}
  \mkern 6mu
  \raise 1pt \hbox {.}
  \mkern 6mu
  \raise -3pt \hbox {.}
}

\newcommand{\bq}{\begin{equation}} 
\newcommand{\eq}{\end{equation}}

\newcommand{\beq}{\begin{equation}}
\newcommand{\eeq}{\end{equation}}
\newcommand{\beqa}{\begin{eqnarray}}
\newcommand{\eeqa}{\end{eqnarray}}

\long\def\check#1{}
\long\def\hide#1{}

\def\HST{{\it HST }}
\newcommand{\Gaia}{{\it Gaia }}

\newcommand{\adder}{$0.029 \pm 0.037 \ $}
\newcommand{\mult}{$1.027 \pm 0.035 \ $}
\newcommand{\nceph}{224\ }

\begin{document} 

\title{The Accuracy of the Hubble Constant Measurement Verified through Cepheid Amplitudes}


\author{Adam G.~Riess}
\affiliation{Space Telescope Science Institute, 3700 San Martin Drive, Baltimore, MD 21218, USA}
\affiliation{Department of Physics and Astronomy, Johns Hopkins University, Baltimore, MD 21218, USA}

\author{Wenlong Yuan}
\affiliation{Department of Physics and Astronomy, Johns Hopkins University, Baltimore, MD 21218, USA}

\author{Stefano Casertano}
\affiliation{Space Telescope Science Institute, 3700 San Martin Drive, Baltimore, MD 21218, USA}
\affiliation{Department of Physics and Astronomy, Johns Hopkins University, Baltimore, MD 21218, USA}

\author{Lucas M.~Macri}
\affiliation{Texas A\&M University, Department of Physics and Astronomy, College Station, TX 77845, USA}

\author{Dan Scolnic}
\affiliation{Duke University, Department of Physics, Durham, NC 27708, USA}

\begin{abstract} 

The accuracy of the Hubble constant measured with extragalactic Cepheids depends on robust photometry and background estimation in the presence of stellar crowding. The conventional approach accounts for crowding by sampling backgrounds near Cepheids and assuming they match those at their positions. We show a direct consequence of crowding by unresolved sources at Cepheid sites is a reduction in the fractional amplitudes of their light curves. We use a simple analytical expression to infer crowding directly from the light curve amplitudes of $>200$~Cepheids in 3 SNe~Ia hosts and NGC~4258 as observed by \HST -- the first near-infrared amplitudes measured beyond the Magellanic Clouds. Where local crowding is minimal, we find near-infrared amplitudes match Milky Way Cepheids at the same periods.  At greater stellar densities we find that the empirically measured amplitudes match the values predicted (with no free parameters) from crowding assessed in the conventional way from local regions, confirming their accuracy for estimating the background at the Cepheid locations. Extragalactic Cepheid amplitudes would need to be $\sim$20\% smaller than measured to indicate additional, unrecognized crowding as a primary source of the present discrepancy in H$_0$. Rather we find the amplitude data constrains a systematic mis-estimate of Cepheid backgrounds to be \adder mag, more than $5\times$ smaller than the size of the present $\sim$0.2 mag tension in H$_0$. We conclude that systematic errors in Cepheid backgrounds do not provide a plausible resolution to the Hubble tension.

\end{abstract} 


\section{Introduction}

A leading approach to measure the Hubble constant (H$_0$) locally uses \HST observations of Cepheid variables in the hosts of recent, nearby Type Ia supernovae (SNe~Ia) to build a 3-rung distance ladder \citep[][hereafter R16]{Riess:2016}.  Cepheids are favored as primary distance indicators because they are very luminous ($M_V \approx -6$ mag), extremely precise \citep[3\% in distance per source,][hereafter R19]{Riess:2019b}, easy to identify due to their periodicity \citep{Leavitt:1912}, and well understood as a consequence of stellar pulsation theory \citep{Eddington:1917}.   They are also the best calibrated tool for this role when a consistent photometric system is used along the distance ladder. \HST WFC3/UVIS and IR have been used by the SH0ES Team (R19) to measure Cepheids in SN hosts and for three independent sources of geometric distance calibration of their luminosities: the megamaser host NGC 4258, the Milky Way (and its parallaxes), and
the LMC (via detached eclipsing binaries). Near-infrared (NIR) observations are particularly valuable and are employed to overcome the twin pitfalls of metallicity and dust which limited the first-generation measurements made in the optical \citep{freedman01,sandage06}.

However, all long-range distance indicators have shortcomings.  For measurements of Cepheids at $D \ge 10$ Mpc, even with their great luminosity and the resolution of {\it HST}, it is often not possible to separate Cepheids from their stellar crowds.  Unlike SNe which fade away to offer a clear view of what lies in their midst, Cepheids modulate their brightness but do not vanish.  It is therefore necessary to {\it statistically} estimate the background flux using nearby regions before we can assess the true flux of a Cepheid.   Hereafter we will refer to the superposition of stellar flux on a Cepheid as ``crowding''.
The fact that these backgrounds are not smooth, but rather composed of unresolved point sources, adds noise to a statistical estimate of the background and hence that of the Cepheid flux.   As \citet{Freedman:2019} noted, ``Possibly the most significant challenge for Cepheid measurements beyond 20 Mpc is crowding and blending from redder (RGB and AGB) disk stars, particularly for near-infrared H-band measurements of Cepheids.'' 

 This challenge is not unique to Cepheid photometry.   Since the advent of CCDs and the desire to measure stellar photometry in dense fields, software tools to measure ``crowded-field photometry'' have used knowledge of the point spread function (PSF) to simultaneously constrain the positions and fluxes of overlapping stars and a more uniform background level of unresolved fluctuations \citep{Stetson:1987,Mateo:1989}.    A drawback of such measurements is that they necessarily rely on the assumption that the background flux {\it near Cepheids} has the same mean as that at the positions of Cepheids.  This would seem a fair assumption because our line of sight to a Cepheid, which determines which stars will be superimposed on the Cepheid, is inherently random.  However, the assumption could fail in the presence of stars {\it physically associated} with the Cepheid and could become important for the distance ladder if the associated flux is not resolved at the distance of SN Ia hosts, but is resolved at the distance Cepheids are geometrically calibrated, a physical scale of 400 Astronomical Units to a few parsecs for Cepheids measured at $D \ge 10$ Mpc and calibrated with Milky Way parallaxes.  A wide-binary or host cluster could be the source of such flux.  \citet{Mochejska:1999} claimed a strong bias from this source in optical Cepheid photometry from the \HST Key Project (KP) and the low-resolution WFPC2 based on binning ground-based images of M31.  However, \citet{Ferrarese:2000} demonstrated the effect was quite negligible in the KP data after simulating {\it the selection} of extragalactic Cepheids using artificial star tests.  These tests account for how the underlying sky is brightened and include the impact on the measured PSF and Cepheid light curve, both metrics for selecting Cepheids.  \citet{Ferrarese:2000} further cited a factor which they stated was ``not easily quantifiable''-- such contamination would decrease the amplitude of the Cepheid light curve, reducing its likelihood of being selected.  \citet{Anderson:2018} \citep[see also][]{Senchyna:2015} used \HST imaging of M31 to resolve the clusters near Cepheids in the bands used by SH0ES (R19).  They found that the associated flux from clusters can be substantial, with a mean of 0.3 mag, but the fraction of Cepheids in such clusters (and close enough to their centers to be unresolved at the distances of SN Ia hosts) is very low, $\sim$ 2.5\%, so the  resulting bias on H$_0$ from the product of the two is $<$0.01 mag.  The low fraction of Cepheids seen in clusters is because the cluster dispersement timescale is a factor of $\sim$ 5 shorter on average than Cepheid ages \citep{Anderson:2018}.  This is reassuring for the goal of reaching a 1\% determination of H$_0$, but does depend on the plausible assumption that the fraction of Cepheids in clusters in M31 is similar to those of the large spirals of SN Ia hosts used to build the distance ladder.  
 
   In light of the present discrepancy between the locally-measured value of H$_0$ and the value inferred from the Early Universe in concert with the cosmological model \citep[see][for a review]{Riess:2019c} and the possibility it raises of new physics, it is necessary to subject all aspects of these measurements to increasingly higher levels of scrutiny.  Therefore, it is important to identify a {\it direct} measure of the crowding of Cepheids in SN Ia hosts used for the determination of H$_0$ which is also independent of resolution and hence distance.

	Here we present a new method that {\it directly} tests the accuracy of crowded-field Cepheid photometry and the key assumption that superimposed flux due to crowding can be accurately estimated from its annular vicinity.  Crowding decreases the amplitude of a Cepheid measured {\it in magnitudes}, due to its greater fractional contribution at minimum vs. maximum light.  Thus the observed amplitude of a Cepheid provides a direct test of this assumption.  In \S 2 we describe crowding and derive a simple mathematical function that relates the apparent amplitude of a Cepheid to its crowding with no free parameters.  In \S 3 we calibrate the amplitudes of Milky Way Cepheids as a function of their period to use as a benchmark to compare to extragalactic Cepheids in the absence of crowding and in \S 4 we present  the first sample of distant NIR extragalactic Cepheid light curve amplitudes. By combining these amplitude-crowding and amplitude-period relations we show that the amplitudes of extragalactic Cepheids match the values expected if their locally-measured crowding is the same as that at the position of the Cepheids, testing a crucial assumption of the distance ladder. This eliminates a possible systematic error at $5\sigma$ confidence for explaining the present discrepancy in H$_0$.

\section{Amplitudes and Crowding}

\subsection{Crowding} 

The SH0ES program identifies Cepheids and measures their periods using 11 to 15 epochs of optical {\it HST} imaging in the hosts of recent SNe~Ia and follows these identifications with NIR Cepheid photometry \citep{riess09a}.  Time-sampled imaging centered near the visual band ($\sim 0.5\mu$m) coupled with the fine $0.04\arcsec$/pixel sampling of WFC3-UVIS and PSF-fitting provides Cepheid photometry at $D= 20$ to 40 Mpc with minimal crowding measured to be $\sim$ 2\% of the Cepheid flux \citep{Hoffmann:2016}.  However, in the NIR the resolution and pixel sampling is a factor of $\sim$ 3 lower  (i.e., for WFC3-IR in the $F160W$, the band is centered at $\sim 1.5\mu$m and has $0.13\arcsec$/pixel) and thus the background area and number of potential stellar contaminators is an order of magnitude greater than in the optical.  In addition, the dominant sources of Cepheid crowding in the NIR are red giants and thus the contrast with the bluer Cepheids is less relative to optical passbands \citep[see Figure 8 in][for examples]{riess09a}.  As a result, the flux from crowding which is within the resolution element of NIR imaging with \HST can rival the flux of the Cepheid\footnote{Photometry measured by summing the flux in apertures after sky subtraction naturally removes the statistical contribution of crowding because it is included in the mean sky level per-pixel, but this method is otherwise disadvantageous in dense regions because it cannot separate overlapping sources using knowledge of the PSF nor does it use the PSF as weights for the target's pixels which improves signal-to-noise.  The need to explicitly account for crowding in PSF photometry results from full blending of background sources with the target PSF rendering the multi-source model degenerate to the presence of such blended sources.  Therefore the presence of such blended sources is inferred statistically from the local stellar density and measured using artificial stars.}.

To accurately measure Cepheid photometry it is therefore critical to characterize and account for the mean level of crowding, which is most readily achieved by adding and measuring artificial stars of the same brightness as each Cepheid, placed randomly in its vicinity.  It is advantageous and most compact to quantify the crowding offset and its dispersion in the units of the difference between the input and output magnitude of the artificial stars because these values have been empirically shown to be distributed as log-normal in flux or Gaussian in magnitudes out to a few standard deviations due to the distribution of background sources whose numbers decrease with increasing flux \citep{riess09a}.  

In magnitude space, we define the crowding, $ \Delta m $, as the difference
between the true magnitude $ m_0 $, coresponding to flux $ F_0 $,
and the apparent measured magnitude $ m' $, corresponding to the 
blended flux $ F_0+F_1 $.  

A crucial advantage for NIR follow-up of optical discoveries is that the Cepheid position in the NIR scene is fixed by the optical image which constrains the fit and lowers the  uncertainty in the NIR measurement.  A further refinement of the analysis for each Cepheid comes from measuring the displacement of its detected position in the NIR image from its optically determined position, a measure of the {\it specific} degree of blending.   The artificial star trials used to characterize the crowding are selected from those with similar displacements \citep{riess09a}.  Lastly, the estimate of the input magnitude for the artificial stars is derived independent of the measured magnitude of the Cepheid by using an estimate derived from its period and the crowding-corrected period-magnitude relation of its host brethren.  This is necessarily an iterative approach as each loop produces a more accurate determination of the unbiased Period-Luminosity relation.

\subsection{Amplitudes}

In Figure 1 we illustrate how crowding alters the measurement of the amplitude of a Cepheid. We can derive a simple relation for the measured amplitude of a symmetric light curve in magnitude space  in the presence of background light.  The approximation is valid for a generic light curve shape, but for definiteness, we consider a sinusoidal curve - which is a good approximation to a Cepehid light curve in the $F160W$ band.
 For a variable flux source  
\begin{equation}
F(t) = F_0 * (1 + \delta(t))
\end{equation} where the mean flux $ \langle F \rangle = F_0 $ and 
we assume for simplicity 
$ \delta = \alpha * \sin (2\pi t/P) $, where $ P $ is the 
period of the variable.  Let's define $ \alpha $ as the maximum 
value of $ \delta(t) $; because of symmetry, the minimum 
value is $ -\alpha $.  
The fractional flux amplitude $ a $ is then:
\begin{equation}
a \equiv (\max(F) - \min(F)) / \langle F \rangle  = \delta_{max} - \delta_{min} = 2 \alpha.
\end{equation}
Now assume that the same source is blended with another source 
of constant flux $ F_1 $.  The apparent light curve will then be:
\begin{equation}
F' = F_0* (1 + \delta(t)) + F_1
\end{equation}
and the apparent fractional flux amplitude $ a' $ will be:
\begin{equation}
a' = (\max(F')-\min(F')) /  \langle F' \rangle = a * (F_0/(F_0+F_1))
\end{equation}
so that the flux amplitude will be compressed by a factor $ F_0 / (F_0 + F_1) $. From the definition of $\Delta m$,

\begin{equation}
\label{eq:redfac}
F_0 / (F_0 + F_1) = 10^{-0.4*m_0} / 10^{-0.4*m'} = 10^{-0.4*\Delta m}
\end{equation}

\noindent while from the definition of the amplitude in magnitude space, $A$ (the difference between the minimum and maximum magnitude) is

\begin{equation}
A = 2.5 \log_{10} [(1+\alpha)/(1-\alpha)] \sim 1.0857 a + O(\alpha^3)
\end{equation}
where we have used the linear approximation for 
$ \log[(1+x)/(1-x)] \approx 2*x + O(x^3) $, and the factor 
$ 1.0857 \approx 2.5 / \log(10) $ converts from magnitudes to natural 
logarithms.

When the source is blended, the amplitude in magnitude, $ A' $, will similarly
follow:
\begin{equation}
A' \approx 1.0857 a' + O(\alpha^3) .
\end{equation} 

From equations 4 and 5, we obtain
to second order in the amplitude the ratio of the apparent to the true amplitude in magnitude space:
\begin{equation}
A' / A = 10^{-0.4*\Delta m}.
\end{equation}
The approximation is better than 0.0005 mag in $A'$ for $0 < \Delta m < 2$ as we show for two example light curves, a sinusoid and a sawtooth-like, in Figure 2.

As equation 8 shows, in the low-crowding case $\Delta m \sim 0$, the observed amplitude $A'$ is the same as the true, $A$, as expected.  For very large crowding $\Delta m \gg 1$, $A' \sim 0$, also as expected.  Equation (8) shows that the apparent amplitude of a Cepheid, $A'$, provides a direct measure of its crowding as long as the true amplitude $A$ can be estimated {\it a priori}. Thus, measuring the apparent amplitude offers the means to test the accuracy of the crowding measured from nearby regions.  To make use of this relation we first need to determine the true amplitudes of Cepheid variables.

\section{Calibrating the Cepheid amplitude-period relation in the Milky Way}

We use Milky Way Cepheids to calibrate the relation between Cepheid amplitudes and periods in the absence of crowding --- including the statistical contribution of close unresolved binaries. These will provide the values of $A$ in equation (8) to compare to the observed amplitudes.

It is well known that the amplitudes of Cepheids vary in a somewhat predictable way with period following the ``Hertzsprung Progression'' \citep{Hertzsprung:1926}, though they exhibit a sizable range at a given period. In Figure 3 and Table 1 we show the amplitudes of 56 Milky Way Cepheids with $P > 10$ days as a function of their periods.  This is the range of periods relevant for the detection of the extragalactic Cepheids.  This sample contains most of all known in the Milky Way and all with readily available $V$ and $H$-band light curves which could be used to determine accurate values of their visual and $H$-band amplitudes, $A^V$ and $A^H$ respectively. The sources of light curves are \citet{laney92}, \citet{monson11}, and \citet{Riess:2018b}.  

The $H$-band amplitude, $A^H$, has a mean value of $\sim 0.4$~mag in this period range but with a rather large dispersion of $\sim 0.1$~mag.  The visual band amplitude, $ A^V $, has a mean value of $\sim 1.0$~mag and a dispersion of $\sim 0.2$~mag.  However, their ratio, $ \frac {A^H} {A^V} $, has a factor of three times lower variation at a given period than $A^H$.  This may be expected because the same physical pulsation produces both amplitudes so the ratio has little variation.  In Figure 3 we see a fairly simple trend of the amplitude ratio $A^H/A^V$ with period which can be fit to good accuracy by a quadratic.  A linear trend for the amplitude ratio with period does nearly as well,  although it does not capture a slight flattening of the ratio at long periods.  

Thus we can predict the H-band amplitude for an extragalactic Cepheid on the basis of its period, and if available, its visual amplitude with good precision.  We recast equation (8) to relate the $V$-to-$H$ amplitude ratio in the presence of crowding (i.e., the primed amplitudes) to the uncrowded (i.e., Milky Way or abbreviated ``MW'' and unprimed) values as 

\begin{equation}
 \left ( \frac {A'^{H}} {A'^{V}}\right ) / \left ( \frac{A^{H,MW}} {A^{V,MW}} \right )=  10^{-0.4*(\Delta m_{H}-\Delta m_{V})}
\end{equation}

Finally, to compare Milky Way Cepheids in $V$ and $H$-bandpasses to extragalactic Cepheids (which are measured in similar bandpasses with overlapping wavelengths: $F350LP$, $F555W$ and $F160W$) we derived and applied the following transformations to the Milky Way amplitudes:

\begin{equation}
A^{F350LP}=A^V/(0.984+0.296*(\log P-1.5))
\end{equation}

and

\begin{equation}
A^{F160W}=1.015*A^H.
\end{equation}

We derived the transformation between $F350LP$ and $V$ from observations of the same Cepheids in both $F350LP$ and $F555W$ in NGC 5584 \citep{Hoffmann:2016}.  We derived transformations between ground and {\it HST} equivalent bandpasses using a subset of Cepheids with multicolor data $V,I,J,H$ resulting in equation (11) and $A^V=1.04*A^{F555W}$. 

As the transformations show, the amplitudes differ by only a few percent so these corrections have minimal impact on the final results but are included for accuracy.  
It may be worth noting that the transformations are needed because the empirically-measured amplitudes (in the Milky Way) are in a different photometric system than the amplitudes measured in SN Ia host galaxies.  If we did not correct for band-dependent ratios, we could under- or over-estimate the predicted unblended amplitudes, leading to an imprecise result.  For simplicity, hereafter we will refer to the amplitudes whether originally measured in $F350LP$, $F555W$ or $V$ but now transformed to the $F350LP$ system via these relations as $A^V$ (primed and unprimed) and for amplitudes measured in $F160W$ or $H$ but now on the $F160W$ system via equation (11) as $A^H$ (primed and unprimed).

Thus we characterize the amplitude ratio for Milky Way Cepheids of a given period (and as shown in Figure 3) as
\begin{equation}
 \frac {A^{H,MW}} {A^{V,MW}} =0.22+0.53 (\log P-1)-0.31 (\log P-1)^2
\end{equation}
with a dispersion of 0.035.  We now proceed to use this relation to compare to the measured amplitude ratios of extragalactic Cepheids of known periods in the presence of a range of crowding.

\section{NIR Amplitudes of  Extragalactic Cepheids vs. Local Crowding}

\subsection{Extragalactic Amplitude Measurements} 

We used multi-epoch NIR imaging with WFC3, designed to find Mira variables, to measure light curves and amplitudes of optically-identified Cepheids.
Such observations are available for SN Ia hosts NGC 5643, NGC 1559, and NGC 2525 and for NGC 4258, the megamaser host used for the geometric calibration of Cepheid luminosities.

We obtained multi-epoch NIR Cepheid photometry using the DAOphot and ALLSTAR packages \citep{Stetson:1987,Stetson:1990}.  Example light curves and their host regions are shown for Cepheids identified optically in $F350LP$ in NGC 1559, the most distant host in our sample at $D \sim 20$ Mpc, in Figure 4.    These display a range of apparent amplitudes and local stellar densities.  

   To help measure the NIR amplitudes we use optical light curves, which have greater leverage to constrain the period and phase of each Cepheid.  Thus we limit the number of parameters used to fit the NIR light curves to two: the mean magnitude and amplitude of a sinusoidal function.  We include a mean phase offset between the NIR and visual light curves of 0.3 \citep[as derived by][from the mean of many Milky Way light curves]{soszynski05} and propagate an uncertainty of $\sigma=0.04$ mag which translates into a 10\% uncertainty in $A'^H$.   Example light curve fits are shown in Figure 4.

In Table 3 we give the measured visual and NIR amplitudes of \nceph Cepheids across 4 hosts which provided good-quality light curve fits, their locally derived crowding corrections in the $V$ and $H$ bands, and periods.  The crowding corrections, $\Delta m_H$, were determined from artificial star injection and retrieval following the same procedure and software used by R16 and described in \S 2.  

Artificial star tests  demonstrate that Cepheid measurements in visual bands ($F555W$ or $F350LP$) at these distances with WFC3 suffer from very little crowding \citep{Hoffmann:2016}, and we find mean values of $\Delta m_{V}$ of 0.009, 0.027, 0.020, and 0.020 mag for NGC 5643, NGC 1559, NGC 2525 and NGC 4258, respectively.  The first 3 were obtained with $F350LP$ and NGC 4258 with $F555W$ imaging.  From equation (8) we thus expect the visual amplitudes to be thus reduced from their true values by only 1-3 \%.  As discussed in \S 2, due to increased pixel size, larger PSF FWHM and reduced contrast, the $F160W$ measurements have greater values of $\Delta m_{H}$.  {\it We have extended the Cepheid sample here to include objects found in high surface brightness regions and with much greater crowding than the limit typically used to measure H$_0$ such as in R16.}  The purpose of this extension is to increase the sensitivity of our sample for testing the relation between Cepheid amplitude and crowding.   The median $\Delta m_{H}$ for the ``low crowding'' sample used to measure $H_0$ (R16) is 0.38 mag and for the ``high crowding'' sample not typically used to measure $H_0$ but added here to increase our sensitivity to study crowding is 0.97 mag as indicated in Figure 5.  In some cases for the high crowding sample $\Delta m_{H} \sim 2.0$ mag, i.e., 5 times the Cepheid for the maximally-crowded case.  

\subsection{The observed crowding-amplitude relation and local crowding}

Following equation (9), the crowding term, $\Delta m_{H} -  \Delta m_{V}$, fully specifies the relation between the extragalactic (crowded) and Milky Way (uncrowded) amplitude ratio with no free parameters so that,

\begin{equation}
 \begin{split}
\left (\frac {A'^{H}} {A'^{V}}\right ) = \left (\frac{A^{H,MW}} {A^{V,MW}}\right ) 10^{-0.4*(\Delta m_{H}-\Delta m_{V})}= \\
(0.22+0.53 (\log P-1)-0.31 (\log P-1)^2) 10^{-0.4*(\Delta m_{H}-\Delta m_{V})} 
 \end{split}
\end{equation}
We substitute the polynomial expression relating ${A^{H,MW}}/{A^{V,MW}}$ to period in equation (12) allowing us to predict the measured amplitude ratios of the extragalactic Cepheids solely from their periods and local crowding and thus without any free parameters on the right hand side of equation (13).

In Figure 5 we show the observed relation between the local crowding $\Delta m_{H}  -  \Delta m_{V}$ and the amplitude ratio of extragalactic to Milky Way Cepheids (we note as discussed earlier, the term $(\Delta m_{H}-\Delta m_{V}) \sim \Delta m_{H}$ to good approximation).

Equation (13) predicts and the measurements in Figure 5 show that at $(\Delta m_{H}-\Delta m_{V}) \sim 0$, Milky Way and extragalactic amplitude ratios are consistent.  As crowding increases, the amplitude ratio for extragalactic Cepheids decreases and approaches zero.  This is confirmed by comparing the fit of equation (13) and a second order polynomial fit constrained only by the data in Figure 5.   We verify that this reduction in ratio with enhanced local crowding is due to the reduction in $A^{H}$ (and not an increase in $A^{V}$) by examining the composite light curves of the $F160W$ light curve points for Cepheids with high and low $\Delta m_{H}$ in Figure 6.  

Next we define a standard $\chi^2$ statistic to assess the goodness of fit of equation (13):

\begin{equation}
 \chi^2=\sum_{i=1}^{n} \left (\frac {A'^{H}} {A'^{V}} - \frac{A^{H,MW}} {A^{V,MW}}10^{-0.4*(\Delta m_{H}-\Delta m_{V})} \right )^2  \sigma_i^{-2}
\end{equation}

The model error in the amplitude ratio, $\sigma_i$, is the quadrature sum of the amplitude measurement error, the dispersion of the local crowding estimate times the amplitude ratio, the error due to phase variations from the fixed offset and the dispersion in the Milky Way amplitude-period relation, with means of 0.079, 0.071, 0.04 and 0.035, respectively.  
  The mean of $\sigma_i$ for all Cepheids is 0.123 with a full range of 0.062 to 0.175; individual values are given in Table 3.  The $\chi^2$ is 229 for \nceph degrees of freedom, a ratio of 1.02, showing that the parameter-free model is a good fit to the data.  
  
\subsection{Constraining departures from the crowding-amplitude relation}

We can use the amplitude data to constrain the size of a systematic over- or under-estimate of Cepheid crowding allowing for a difference, $\gamma$, between the crowding at the position of the Cepheid $\Delta m_H + \gamma$ and the value inferred nearby, $\Delta m_H$,

\begin{equation}
 \chi^2(\gamma)=\sum_{i=1}^{n} \left ( \frac {A'^{H}} {A'^{V}} - \frac{A^{H,MW}} {A^{V,MW}}10^{-0.4*(\Delta m_{H}-\Delta m_{V}+\gamma)} \right)^2 \sigma_i^{-2}
\end{equation}

Minimizing $\chi^2$ yields $\gamma=$ \adder mag as shown in Figure 7.  Thus, to a precision of $\sigma \leq 0.04$ mag, the mean NIR crowding at the position of the Cepheid matches that inferred from nearby regions.    This rules out at $5\sigma$ a systematic mis-estimate of Cepheid crowding as a primary cause of the well-known 0.2 mag discrepancy between local and cosmological estimates of H$_0$ \citep{Riess:2019c}; see Section 5  for more details.  Following \citet{Anderson:2018} we can also place a constraint on the fraction of Cepheids in clusters as less than 25\% at $>$ 95\% confidence.

The unique value of the amplitude data is as a measure of the crowding of Cepheids.  However, as an alternative to this analysis we can test whether extragalactic Cepheid amplitudes match those of Milky Way Cepheids by allowing for a re-scaling of the Milky Way Cepheids, $\alpha$, to best match the extragalactic sample, accounting for crowding locally, by minimizing the following statistic,

\begin{equation}
 \chi^2(\alpha)=\sum_{i=1}^{n} \left ( \frac {A'^{H}} {A'^{V}} - \alpha \left \{\frac{A^{H,MW}} {A^{V,MW}}10^{-0.4*(\Delta m_{H}-\Delta m_{V})}\right \} \right )^2 \sigma_i^{-2}
\end{equation}

We find a best-fit value of $\alpha=$ \mult shown in Figure 7.  That is, the amplitudes ratios of Milky Way and extragalactic Cepheids at the same periods agree to $\leq$ 4\% precision.
 
\section{Discussion}

There are presently three routes for geometrically calibrating the luminosity of Cepheids empirically and in a single photometric system to better than 2\% precision: with masers in NGC 4258, detached-eclipsing binaries in the LMC and parallaxes in the Milky Way.  In the context of crowding as a source of systematic uncertainty, NGC 4258 and specifically its inner region offers the ``advantage'' that its Cepheids suffer similar crowding as those in SN Ia hosts.  Therefore a systematic over- or under-estimate of Cepheid crowding would be expected to cancel in the measurements of Cepheids along the distance ladder.  At present all three routes offer similar precision and more than enough to verify the present Hubble tension, so the route via NGC 4258 offers valuable confirmation in the presence of crowding.
The amplitude measurements presented here allow us to rule out a mis-estimate of Cepheid crowding large enough to explain the tension at $\sim 5\sigma$ confidence level.  A restatement of this result in terms of what is measured is that extragalactic Cepheid amplitudes would need to be $\sim$20\% smaller than empirically measured to indicate additional, unrecognized crowding as a primary source of the present discrepancy in H$_0$.
The results presented here provide the third (after the use of NGC 4258 to calibrate Cepheids and the high-resolution study of Cepheid environments in M31) and perhaps most powerful confirmation that crowding does not compromise the accuracy of H$_0$ measured using extragalactic Cepheids.  

Looking forward, the effort to reach a 1\% measurement of the local value of H$_0$ must increasingly rely on parallax measurements of Cepheids from the ESA \Gaia mission as they offer the only means of reaching this highly sought precision.  The expected precision of \Gaia parallaxes and the photometric homogeneity available from \HST observations of Milky Way Cepheids \citep{Riess:2018b} can produce a Cepheid luminosity calibration with $\sim$ 0.4\% precision, provided it is not otherwise degraded.  Because Milky Way Cepheids do not suffer crowding from our vantage point (and it is difficult to simulate their photometry seen from a perspective of $D>$ 10 Mpc), we need to accurately and precisely account for crowding in the measurements of extragalactic Cepheids. Therefore Cepheid amplitudes provide a valuable means to directly test corrections for crowding measured locally.  However, we note that NIR amplitude measurements are costly in terms of observing time (where a single random phase will suffice for accurate distance measurements) and their signal-to-noise ratio is only $\sim$ 4.  Converting the amplitude constraints to individual constraints on crowding yields $\sigma=0.55$~mag -- not as precise as can be derived using the conventional approach even if they were available for all Cepheids.  As a result, while providing a measure of the crowding at the position of the Cepheid whose uncertainty dominates the scatter of the Period-Luminosity relation, the precision of amplitude measurements is not adequate to provide enhanced precision in the determination of these relations.

Greater resolution would be valuable as well.  The {\it James Webb Space Telescope} will offer 3 times the resolution of \HST in the NIR and thus reduce the crowding at those wavelengths by its square or about an order of magnitude in flux, to a level comparable to that of \HST in the visible.  Differential tests comparing Cepheid measurements between \HST and {\it JWST} at similar wavelengths should provide tests of their photometry with fidelity comparable to those based on amplitudes.  Continued efforts to characterize the stellar populations around Cepheids will also be important to further reduce systematics associated with background subtraction.  With all three tools together we may expect enough precision and cross-checks to keep this issue of crowding from degrading the available precision in the measurement of H$_0$ as measurements approach 1\% precision.

\bigskip

\input mwamp.tex

\begin{table}[h]
\begin{small}
\begin{center}
\vspace{0.4cm}
\begin{tabular}{ccccc}
\multicolumn{5}{c}{{\bf Table 2:} Multi-epoch Observations of hosts in $F160W$ }\\
\hline
\hline

Host  & epochs  & exposure  & \HST programs & $\#$ Cepheids$^*$  \\
  &    & (sec)  & & \\
\hline
NGC 5643 & 9 & 1000 & 15145,15640 & 102 \\
NGC 1559 & 9 & 1000 & 15145 & 78 \\
NGC 2525 & 13 & 900-1100 & 15145,15693 & 15 \\
NGC 4258i & 12 & 2200 & 13445 & 27\\
NGC 4258o & 4 &  1500 & 15640 & 2\\
\hline
\hline
\multicolumn{5}{l}{$^*$ Cepheids with well-fit NIR light curves, $\sigma_{amp} < 0.15$}\\
\end{tabular}
\end{center}
\end{small}
\end{table}

\input examp.tex

\acknowledgements

Support for this work was provided by the National Aeronautics and Space Administration (NASA) through programs GO-14648, 15146 from the Space Telescope Science Institute (STScI), which is operated by AURA, Inc., under NASA contract NAS 5-26555. A.G.R., S.C. and L.M.M. gratefully acknowledge support by the Munich Institute for Astro- and Particle Physics (MIAPP) of the DFG cluster of excellence ``Origin and Structure of the Universe.''

This research is based primarily on observations with the NASA/ESA {\it Hubble Space Telescope}, obtained at STScI, which is operated by AURA, Inc., under NASA contract NAS 5-26555.

The \HST data used in this paper are available as part of the MAST archive which can be accessed at {\tt http://archive.stsci.edu/hst/}.

\include{figs2}

\vfill
\eject

\clearpage
\bibliographystyle{aasjournal} %
\bibliography{bibdesk}
\clearpage
\end{document}

%% file: mwamp.tex
\startlongtable
\begin{deluxetable*}{cccc}
\tabletypesize{\normalsize}
\tablewidth{0pt}
\tablenum{1}
\tablecaption{Amplitude Data for Milky Way Cepheids\label{tb:amp1}}
\tablehead{\colhead{Cepheid} &  \colhead{Period} & \colhead{$A^V$} & \colhead{$A^H$}}  
\startdata
SY-Aur &  10.15  &  0.663  &  0.170  \\
VX-Per &  10.89  &  0.684  &  0.207  \\
Z-Lac &  10.89  &  0.976  &  0.277  \\
SV-Per &  11.13  &  0.881  &  0.266  \\
DR-Vel &  11.20  &  0.728  &  0.274  \\
AA-Gem &  11.30  &  0.602  &  0.205  \\
RX-Aur &  11.62  &  0.675  &  0.235  \\
UU-Mus &  11.64  &  1.082  &  0.298  \\
RY-Cas &  12.14  &  0.970  &  0.235  \\
KK-Cen &  12.18  &  1.022  &  0.270  \\
SS-CMa &  12.35  &  0.982  &  0.219  \\
XY-Car &  12.44  &  0.876  &  0.303  \\
SY-Nor &  12.65  &  0.874  &  0.282  \\
Z-Sct &  12.90  &  1.007  &  0.290  \\
AD-Pup &  13.60  &  1.119  &  0.317  \\
BN-Pup &  13.67  &  1.227  &  0.429  \\
CY-Aur &  13.85  &  0.951  &  0.346  \\
SV-Vul &  14.10  &  1.036  &  0.370  \\
SV-Vel &  14.10  &  1.148  &  0.249  \\
RW-Cas &  14.79  &  1.178  &  0.364  \\
VW-Cen &  15.04  &  1.032  &  0.338  \\
SZ-Cyg &  15.11  &  0.913  &  0.332  \\
XX-Car &  15.71  &  1.263  &  0.381  \\
RW-Cam &  16.42  &  0.845  &  0.347  \\
XZ-Car &  16.65  &  1.033  &  0.361  \\
CD-Cyg &  17.08  &  1.210  &  0.401  \\
CP-Cep &  17.86  &  0.781  &  0.337  \\
YZ-Car &  18.17  &  0.797  &  0.249  \\
VY-Car &  18.90  &  1.076  &  0.390  \\
RU-SCT &  19.70  &  1.108  &  0.392  \\
KX-CYG &  20.05  &  1.171  &  0.373  \\
VX-CYG &  20.14  &  0.988  &  0.386  \\
RY-Sco &  20.32  &  0.823  &  0.284  \\
RZ-Vel &  20.40  &  1.192  &  0.460  \\
V340-Ara &  20.81  &  1.094  &  0.366  \\
WZ-Sgr &  21.85  &  1.094  &  0.404  \\
BM-PER &  22.96  &  1.332  &  0.545  \\
WZ-Car &  23.01  &  1.212  &  0.543  \\
VZ-Pup &  23.17  &  1.309  &  0.490  \\
VZ-Pup &  23.17  &  1.276  &  0.485  \\
SW-Vel &  23.44  &  1.227  &  0.525  \\
X-Pup &  25.97  &  1.308  &  0.542  \\
OT-PER &  26.16  &  0.961  &  0.405  \\
T-Mon &  27.03  &  0.989  &  0.425  \\
RY-Vel &  28.13  &  0.959  &  0.338  \\
KQ-Sco &  28.66  &  0.898  &  0.370  \\
AQ-Pup &  30.12  &  1.183  &  0.484  \\
V0609-CYG &  31.06  &  1.151  &  0.448  \\
V0396-CYG &  33.25  &  0.842  &  0.394  \\
KN-Cen &  34.03  &  1.047  &  0.425  \\
l-Car &  35.55  &  0.697  &  0.323  \\
U-Car &  38.82  &  1.156  &  0.459  \\
RS-Pup &  41.44  &  1.082  &  0.461  \\
V1467-CYG &  48.53  &  1.012  &  0.404  \\
GY-SGE &  51.53  &  0.647  &  0.293  \\
S-Vul &  69.16  &  0.601  &  0.223  \\
\hline
\enddata
\end{deluxetable*}

%% file: examp.tex
\startlongtable
\begin{deluxetable*}{cccccccc}
\tabletypesize{\scriptsize}
\tablewidth{0pt}
\tablenum{3}
\tablecaption{Amplitude and Crowding Data for Extragalactic Cepheids\label{tb:amp2}}
\tablehead{\colhead{ID} &  \colhead{Host} & \colhead{Log P} & \colhead{$\Delta m_V$}   & \colhead{$\Delta m_H$} & \colhead{$A^V$} & \colhead{$A^H$}   & \colhead{$\sigma^*$}} 
\startdata
137184 & n5643 &  1.2775  &  -0.008  &  1.012  &  0.722  &  -0.030  &  0.154  \\
359695 & n5643 &  1.3822  &  0.006  &  0.548  &  1.073  &  0.451  &  0.119  \\
322246 & n5643 &  1.8627  &  0.031  &  0.215  &  0.396  &  0.257  &  0.091  \\
351118 & n5643 &  1.4188  &  -0.025  &  0.531  &  0.995  &  0.270  &  0.147  \\
203310 & n5643 &  1.3248  &  0.016  &  1.134  &  0.771  &  0.275  &  0.132  \\
\hline
\enddata
\tablecomments{$^*$ Combined error, $\sigma_i$ for eq 14-17 including dispersion in MW relation, 0.05 intrinsic scatter and propagated uncertainty in $\Delta m_H$ }
\tablecomments{Full table available upon publication }

\end{deluxetable*}

%% file: figs2.tex
\begin{figure}[ht]
\vspace*{150mm}
\figurenum{1}
\includegraphics{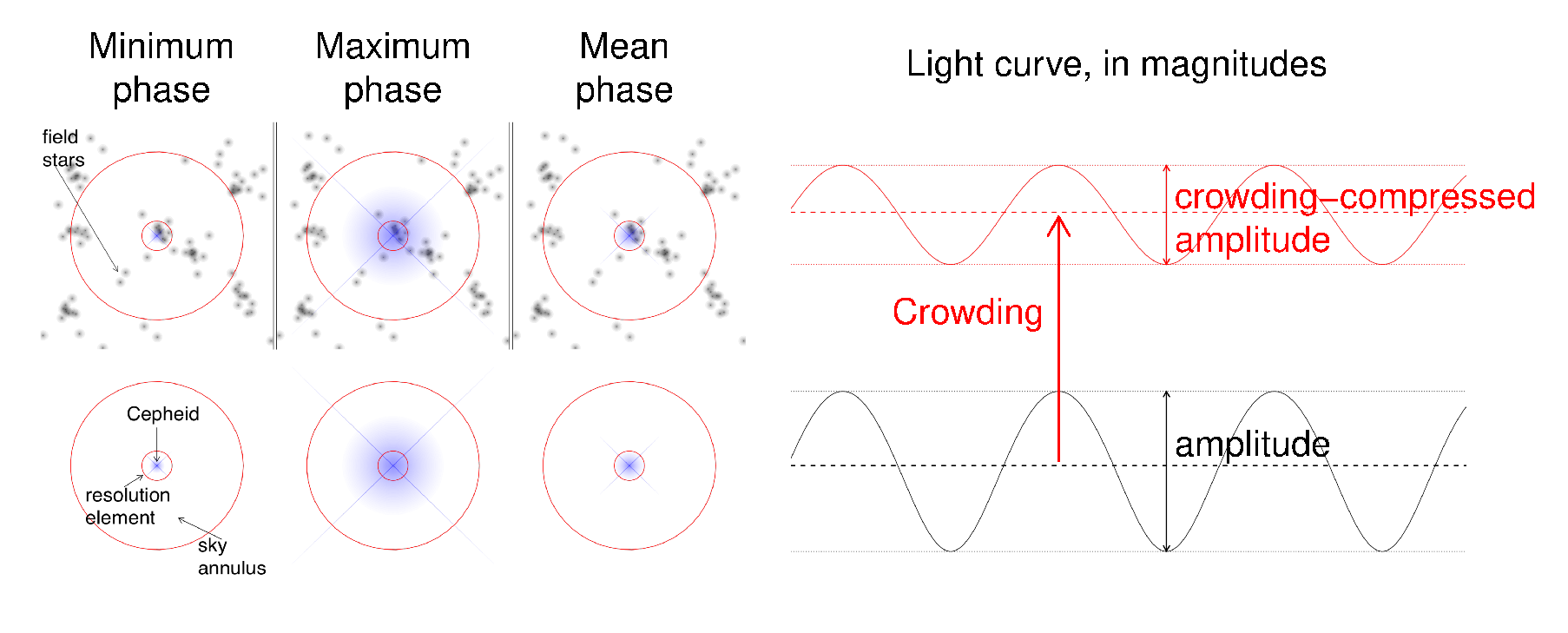}
\caption{\label{fg:outlier}  Illustration of the relation between crowding and apparent amplitude in magnitude space.  Upper left: Crowded case, photometry of a Cepheid includes flux from field stars contained in its resolution element.  The resulting light curve in magnitude space (upper right) is brighter (i.e., crowding) and its amplitude is compressed (because magnitudes measure fractional flux and the fractional contribution of the field stars flux is greater at minimum phase than at maximum phase) Lower panels: Uncrowded case.}
\end{figure}

\begin{figure}[ht]
\vspace*{150mm}
\figurenum{2}
\includegraphics{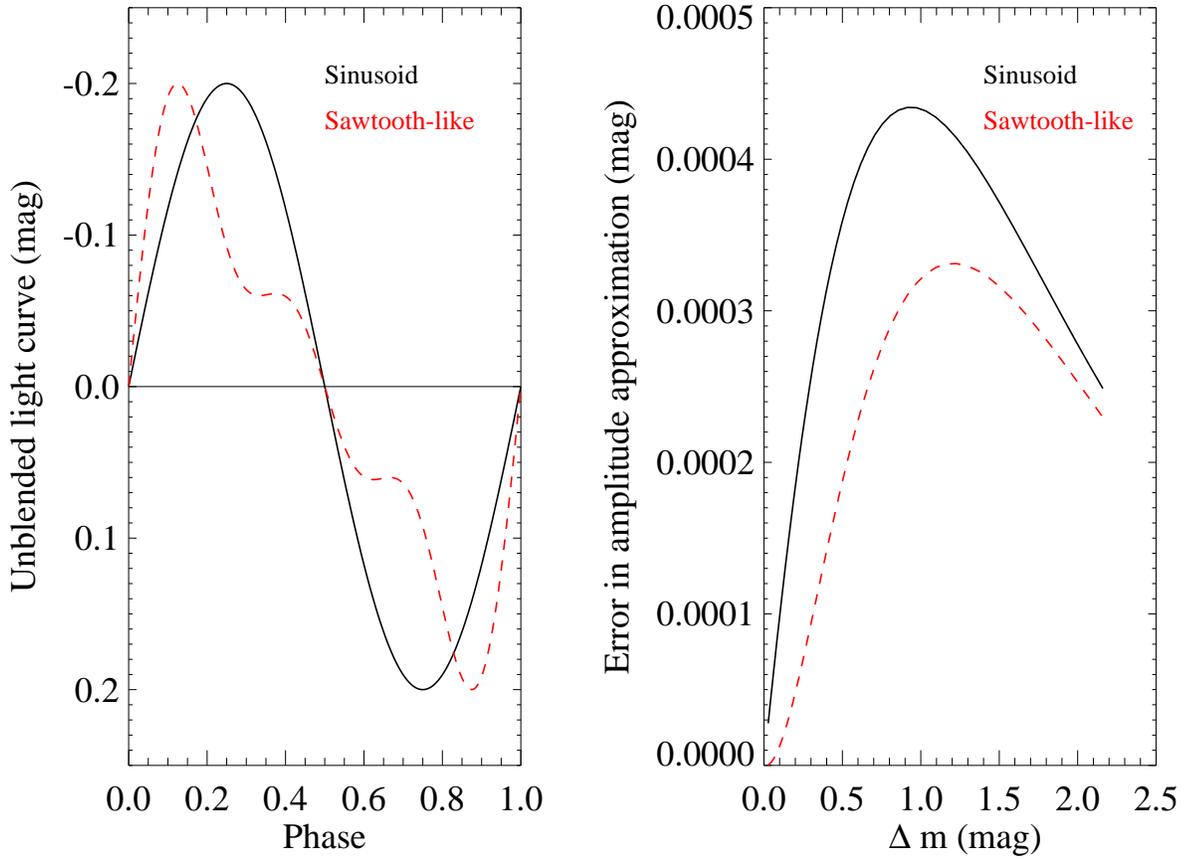}
\caption{\label{fg:outlier}  Numerical vs analytical approximation for computing the
impact of crowding on amplitude.   Two example light curves are given on the left: a sinusoid and a saw-tooth.  The difference between a numerical computation vs. the analytical approximation of equation (8) is shown on the right.  As shown, the analytical approximation is good to 0.2 \%, a far smaller difference than other sources of uncertainty.}
\end{figure}

\begin{figure}[ht]
\vspace*{150mm}
\figurenum{3}
\includegraphics{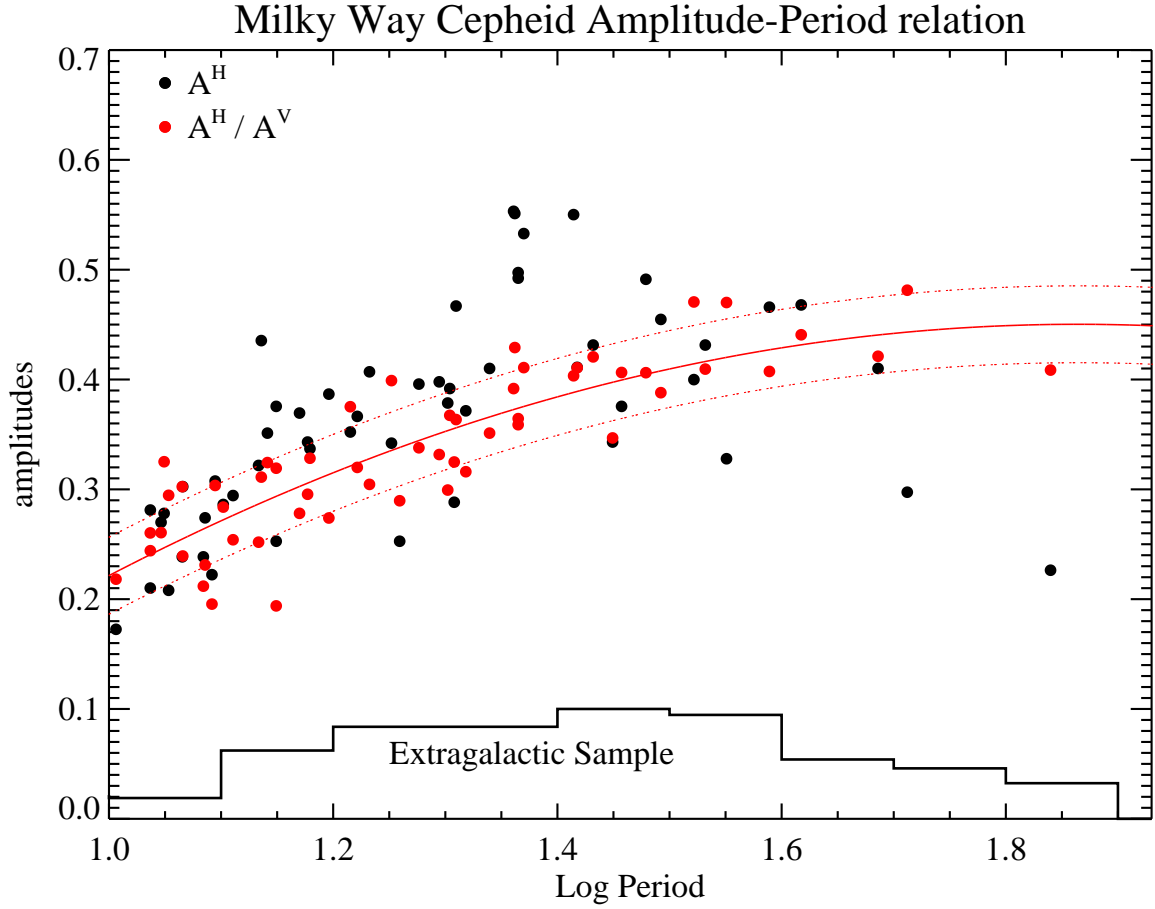}
\caption{\label{fg:outlier}  Relation between light curve amplitude and period from Milky Way Cepheids with $\log P > 1$. $H$-band amplitudes ($A^{H}$) are plotted using black circles, while red symbols denote the amplitude ratio ${A^{H}}/{A^{V}}$.  The red line shows a quadratic fit to this data and is used to model this amplitude ratio as a function of period.  The histogram at the bottom shows the period distribution of extragalactic Cepheids whose amplitudes will be compared to Milky Way Cepheids.}
\end{figure}

\begin{figure}[ht]
\vspace*{150mm}
\figurenum{4}
\includegraphics{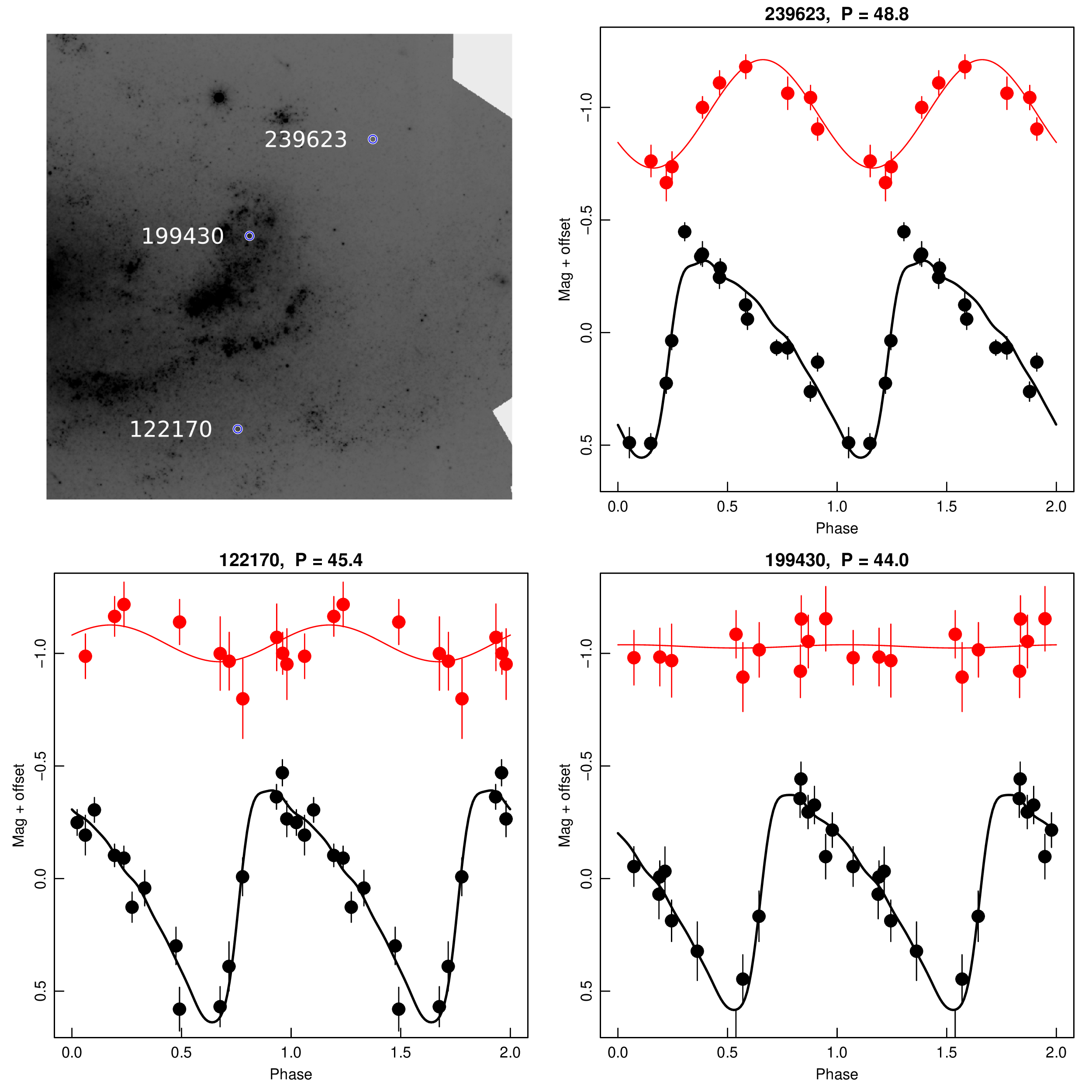}
\caption{\label{fg:outlier}  Three example light curves of Cepheids in NGC 1559, at $D = 20$ Mpc, the most distant in our sample. These showcase amplitudes which are nominal for Milky Way Cepheids (upper right), half as normal (lower left) and minimal (lower right), a progression that statistically follows their local stellar density as shown in the upper left.  The latter example is in a region with higher density than the limit imposed for measurements of $H_0$.   Black points are from $F350LP$ and red from $F160W$.}
\end{figure}

\begin{figure}[ht]
\vspace*{150mm}
\figurenum{5}
\includegraphics{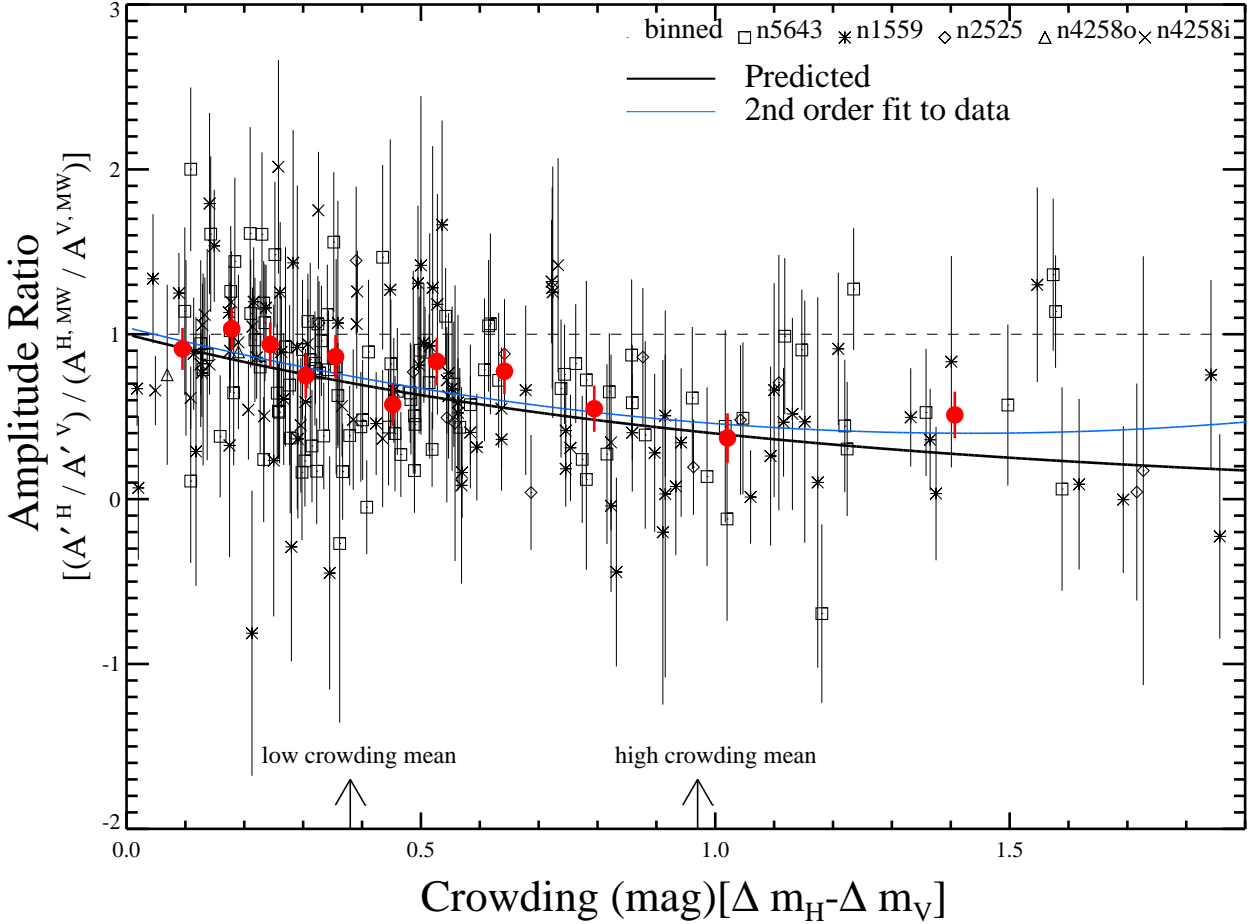}
\caption{\label{fg:outlier}  Amplitude ratios vs crowding of Cepheids for 4 hosts.  The x-axis is the local crowding measured as the difference between the input and recovered magnitudes for artificial stars.  In practice this term is dominated by the NIR crowding term, $\Delta m_{H}$.  The y-axis is the ratio of amplitudes for extragalactic Cepheids compared to Milky Way Cepheids at the same period.  The predicted trend is given by equation (9) with no free parameters and is plotted as the black line.  A quadratic fit to the data (blue line) gives similar results showing that the locally-measured crowding is, on average, a good approximation to the true crowding of the Cepheid. The dashed line is plotted for guidance. The median $\Delta m_{H}$ for the ``low crowding'' sample used to measure $H_0$ (R16) is 0.38 mag and for the ``high crowding'' sample which is not typically used to measure $H_0$ but added here to increase our sensitivity to study crowding is 0.97 mag.}
\end{figure}

\begin{figure}[ht]
\vspace*{150mm}
\figurenum{6}
\includegraphics{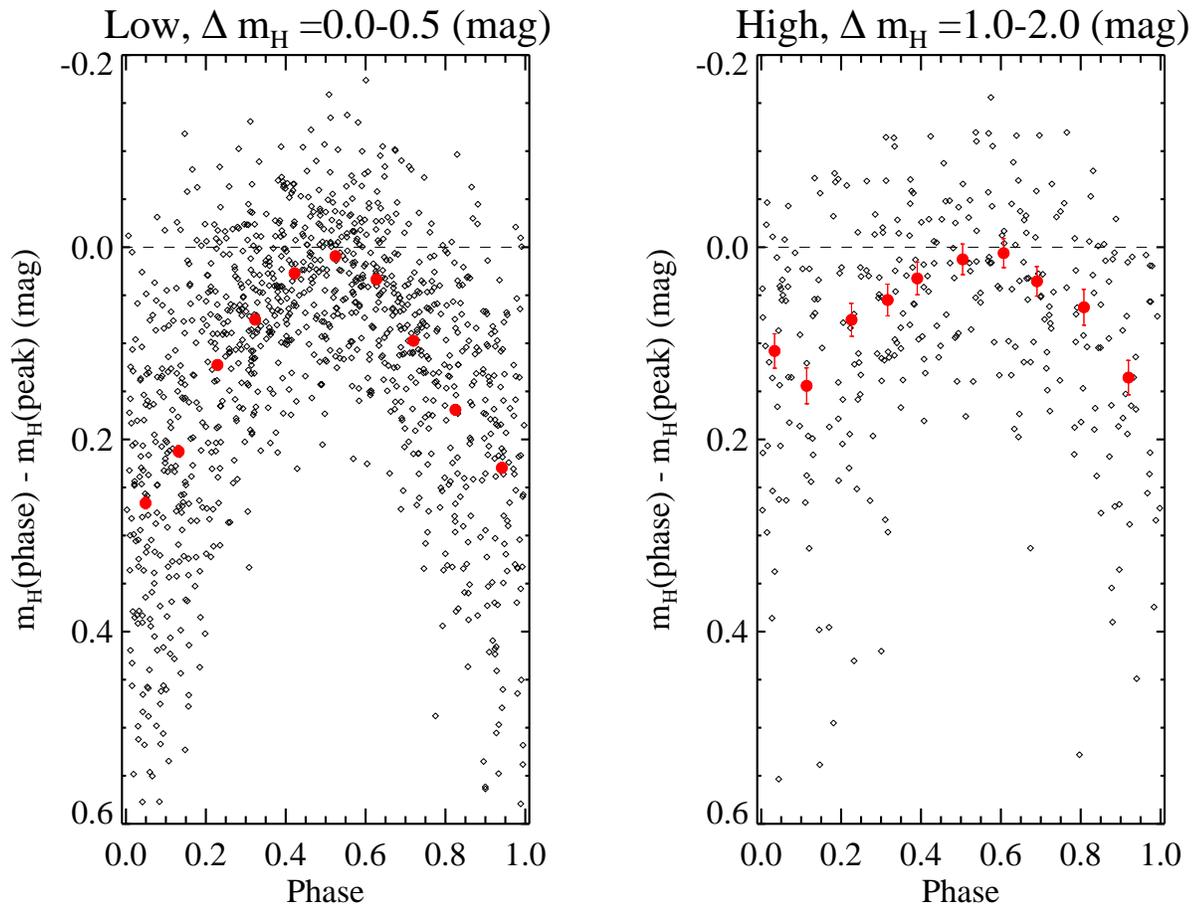}
\caption{\label{fg:outlier}  Composite NIR light curves in low- and high-crowding bins.  Left: low-crowding cases, composite light curve has MW-like amplitude.  These Cepheids are representative of those used for the distance scale.   Right: high-crowding cases, composite light curve amplitude noticeably diminished. Objects with this level of crowding are typically not used for the distance scale.}
\end{figure}

\begin{figure}[ht]
\vspace*{150mm}
\figurenum{7}
\includegraphics{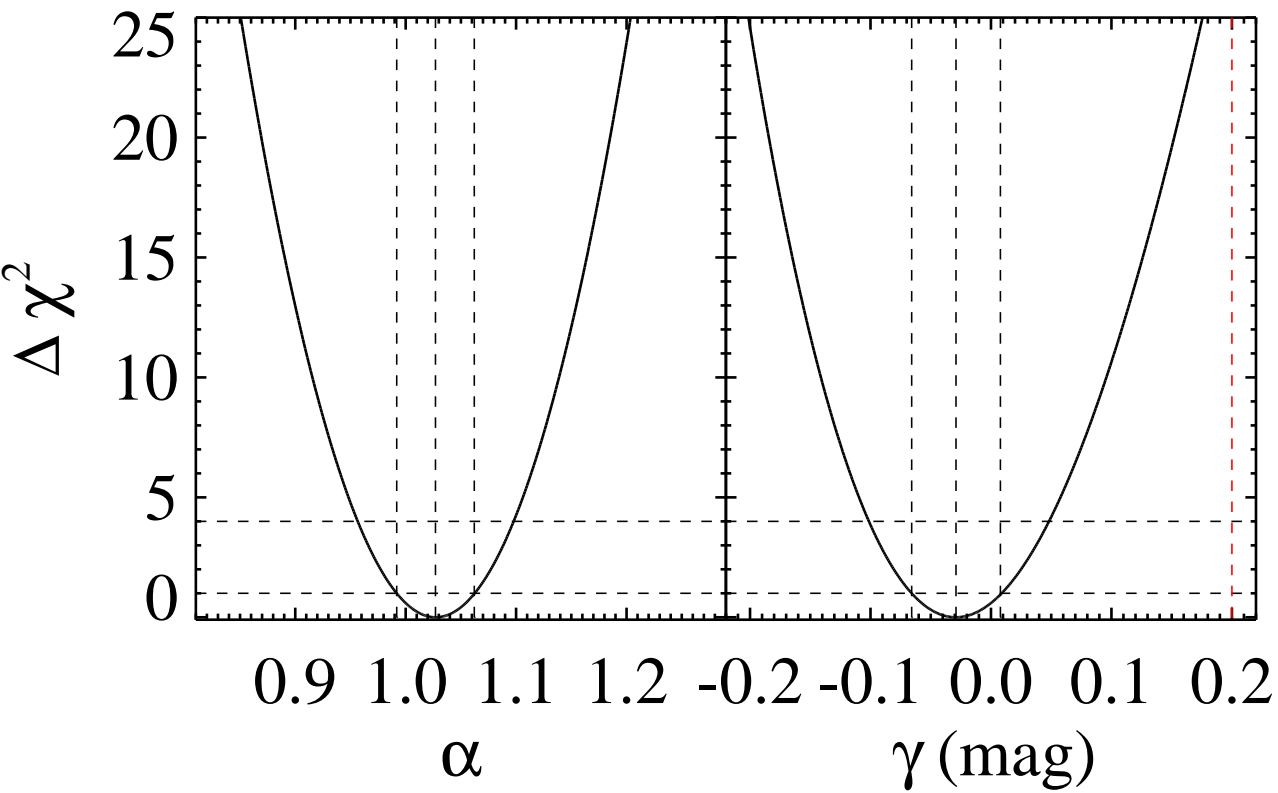}
\caption{\label{fg:outlier}  $\chi^2$ tests of the amplitude and crowding data.  Left:  allowing for a best rescaling, $\alpha$, of the Milky Way Cepheid amplitudes to match extragalactic as in equation (16).  Right: allowing for difference between the crowding at the position of the Cepheid vs a local determination, $\gamma$, as described in equation (15).  Red vertical line shows level needed to solve the Hubble tension.}
\end{figure}